\documentclass[12pt,letterpaper,english]{article}
\usepackage[T1]{fontenc}
\usepackage{babel}
\usepackage{graphics}

\makeatletter

\providecommand{\LyX}{L\kern-.1667em\lower.25em\hbox{Y}\kern-.125emX\@}

\usepackage[T1]{fontenc}


\makeatother
\begin{document}
\baselineskip .3in 

\begin{titlepage}

{\centering \textbf{\Large Failure due to fatigue in fiber bundles
and solids }\Large \par}

\vskip .3in 

{\noindent \centering \textbf{Srutarshi Pradhan} \( ^{(1)} \) and
\textbf{Bikas K. Chakrabarti} \( ^{(2)} \)\par}

{\noindent \centering \textit{Saha Institute of Nuclear Physics},\\
 \textit{1/AF Bidhan Nagar, Kolkata 700 064, India.}\\
 \par}

\noindent \vskip .4in

\noindent \textbf{Abstract}

\noindent \vskip .1in

\noindent We consider first a homogeneous fiber bundle model where
all the fibers have got the same stress threshold (\( \sigma _{c} \))
beyond which all fail simultaneously in absence of noise. At finite
noise, the bundle acquires a fatigue behavior due to the noise-induced
failure probability at any stress \( \sigma  \). We solve this dynamics
of failure analytically and show that the average failure time \( \tau  \)
of the bundle decreases exponentially as \( \sigma \rightarrow \sigma _{c} \)
from below and \( \tau =0 \) for \( \sigma \geq \sigma _{c} \).
We also determine the avalanche size distribution during such failure
and find a power law decay. We compare this fatigue behavior with
that obtained phenomenologically for the nucleation of Griffith cracks.
Next we study numerically the fatigue behavior of random fiber bundles
having simple distributions of individual fiber strengths, at stress
\( \sigma  \) less than the bundle's strength \( \widetilde{\sigma _{c}} \)
(beyond which it fails instantly). The average failure time \( \tau  \)
is again seen to decrease exponentially as \( \sigma \rightarrow \widetilde{\sigma _{c}} \)
from below and the avalanche size distribution shows similar power
law decay. These results are also in broad agreement with experimental
observations on fatigue in solids. We believe, these observations
regarding the failure time are useful for quantum breakdown phenomena
in disordered systems.

\noindent \vskip 1.1in

\noindent \textit{e-mail addresses} :

\noindent \( ^{(1)} \)spradhan@cmp.saha.ernet.in

\noindent \( ^{(2)} \)bikas@cmp.saha.ernet.in 

\noindent \end{titlepage}

\noindent \textbf{I. Introduction }

\noindent \vskip .1in

\noindent If one puts a load or stress (\( \sigma  \)) on a solid
or applies a voltage across an electrical circuit, a strain in the
solid or a current through the circuit develops which grows linearly
(Hooke's law or Ohm's law) with the stress or voltage. If the external
load on the system increases beyond its threshold limit (\( \sigma _{c} \)),
the system fails: stress \( \sigma  \) drops to zero due to fracture
of the solid. The same occurs when the voltage on the network exceeds
its limit and the current drops to zero due to the fuse of the circuit.
Similar failures occur in dielectric materials when the electric field
across the sample exceeds beyond its limit, and dielectric breakdown
sets in. These failures usually nucleate around the defects in the
solid and the failure behavior and its statistics therefore crucially
depends on the disorder or impurity distribution within the sample.
These (quasi-static) failure properties of disordered solids have
been studied extensively in recent years \cite{Books-1}. 

The dynamics of these failures in such systems are quite intriguing
and is being studied very intensively these days. The critical dynamics
of failure and its universality class in the democratic (global load
sharing) fiber bundle model \cite{fiber-static} has been established
very recently \cite{fiber-dynamic}. These dynamics of failure are
intrinsic and induced by the successive stress redistributions due
to the failure of weaker fibers. However, an important kind of dynamical
failure due to fatigue \cite{Lawn-93} occurs in such disordered systems
when the fibers have an effective probability to fail under any stress
\cite{coleman-58}, or as the micro-cracks within the solid grow at
the crack-tips with time due to chemical diffusion in the atmosphere
\cite{Lawn-93}. The system then fails under a stress less than its
normal strength (\( \sigma _{c} \)) and the time of failure (\( \tau  \))
depends on the load applied on the sample: \( \tau \neq 0 \) for
\( \sigma <\sigma _{c} \) and \( \tau \simeq 0 \) for \( \sigma \geq \sigma _{c} \).

Here, we study first a phenomenological theory of crack nucleation,
following Griffith \cite{Lawn-93,Feng-91}, at finite temperature
(\( T \)) and estimate the average failure time \( \tau  \) at any
stress \( \sigma  \) less than \( \sigma _{c} \). We then develop
a simple model of fatigue-failure in a democratic fiber bundle model
containing identical fibers of strength \( \sigma _{c} \) (homogeneous
bundle), where the fibers have a finite noise-induced failure probability.
We have derived analytically the failure time for the bundle as a
function of the applied stress (\( \sigma  \)) and the noise (\( \widetilde{T} \)).
This result for the model is compared with that obtained for the phenomenological
theory of crack nucleation at finite temperature. It is also in broad
agreement with some recent experimental observations on fatigue in
disordered solids \cite{Lawn-93,expt-01}. Next, we derive the avalanche
size distribution in this fixed strength model analytically and find
robust power law decay. The above analytic results have been confirmed
through the numerical studies on the same model. Finally we consider
random fiber bundles with simple, yet nontrivial, distributions of
the fiber strengths. Our numerical results show that for all these
fiber bundles, the average time to failure \( \tau  \) decreases
exponentially as the stress level \( \sigma  \) approaches bundle's
strength \( \widetilde{\sigma _{c}} \) from below and the avalanche
size distributions show similar power law decay. We also discuss the
plausibility of this (noise-induced) failure in other similar situations.
In particular, we consider the validity of our model in quantum breakdown
phenomena \cite{bkc-book}: for example, in dielectric breakdown where
the microscopic failure of the dielectric grains acquire a finite
probability at any electric field due to quantum tunneling. The failure
time and its variation with the strength of the external field in
such a quantum failure can give us an estimate of the tunneling frequencies
involved.

\vskip .2in

\noindent \textbf{II. Time for fracture in the Griffith nucleation
model}

\noindent \vskip .1in

\noindent Griffith in 1920, equating the released elastic energy of
a growing crack inside a solid with the energy of the newly created
crack surfaces, came to a quantitative estimate of the fracture strength
of a solid containing an already existing fixed geometry micro-crack.
Assuming the linear elasticity behavior up to the breaking point of
a brittle solid, the released elastic energy becomes \( E_{el}=(\sigma ^{2}/Y)l_{0}^{3} \)
for a three dimensional elastic solid under stress \( \sigma  \),
modulus of elasticity \( Y \), containing a micro-crack of length
\( l_{0} \). The corresponding surface energy \( E_{s}=\phi l^{2}_{0} \)
where \( \phi  \) denotes the (crack) surface energy density. Using
the concept of energy balance, Griffith equated the differential increment
in the elastic energy \( dE_{el} \) with the corresponding surface
energy increment \( dE_{s} \) as the crack propagates a further length
\( dl \) and got \begin{equation}
\label{july24-1}
\sigma _{c}=\frac{\Omega }{\sqrt{l_{0}}},\Omega =\sqrt{Y\phi }
\end{equation}

\noindent for equilibrium extension of the crack. Here \( \sigma _{c} \)
is the amount of stress for and above which the micro-crack propagates
in no time (or in a small time dependent on the sound velocity) and
causes a macroscopic failure of the sample. 

This quasi-static picture can be extended to fatigue behavior of crack
propagation for \( \sigma <\sigma _{c} \). At any stress \( \sigma  \)
less than \( \sigma _{c} \), the cracks can still nucleate \cite{Feng-91}
for a further extension at any finite temperature \( T \) with a
probability \( \sim \exp [-E/k_{B}T] \) and consequently the sample
fails within a failure time \( \tau  \) given by \begin{equation}
\label{july24-2}
\tau ^{-1}\sim \exp [-E(l_{0})/k_{B}T],
\end{equation}

\noindent where \begin{equation}
\label{july24-3}
E(l_{0})=\phi l^{2}_{0}-\frac{\sigma ^{2}}{Y}l^{3}_{0}
\end{equation}

\noindent is the crack (of length \( l_{0} \)) nucleation energy.
Here \( k_{B} \) is the Boltzman factor. One can therefore express
(2) as \begin{equation}
\label{july24-4}
\tau \sim \exp [A(1-\frac{\sigma ^{2}}{\sigma ^{2}_{c}})],
\end{equation}

\noindent where (the dimensionless parameter) \( A=l^{3}_{0}\sigma ^{2}_{c}/(Yk_{B}T) \)
and \( \sigma _{c} \) is given by (\( 1 \)). This immediately suggests
that the failure time \( \tau  \) grows exponentially for \( \sigma <\sigma _{c} \)
and approaches infinity if the stress \( \sigma  \) is much less
than \( \sigma _{c} \) when the temperature \( T \) is small, whereas
\( \tau  \) becomes vanishingly small as the stress \( \sigma  \)
exceeds \( \sigma _{c} \). 

\noindent \vskip .2in

\noindent \textbf{III. Fatigue in a homogeneous fiber bundle }

\noindent \vskip .1in

\noindent Fatigue in fiber bundle model was first studied by Coleman
in 1958 \cite{coleman-58}. Thermally activated failures of fiber
have recently been considered and approximate fatigue behavior has
been studied \cite{Roux-00}. We consider here a very simple fiber
bundle model with noise-induced activated failure, for which the dynamics
can be analytically solved.

Let us consider a homogeneous bundle of \( N \) fibers under load
\( L(=N\sigma ) \), each having identical failure strength \( \sigma _{c} \).
Without any noise (\( \widetilde{T}=0 \)), the model is trivial:
the bundle does not fail (failure time \( \tau  \) is infinity) for
stress \( \sigma <\sigma _{c} \) and it fails immediately (\( \tau =0 \))
for \( \sigma \geq \sigma _{c} \). We now assume that each such fiber
has a finite probability \( P(\sigma ,\widetilde{T} \)) of failure
at any stress \( \sigma  \) induced by a non-zero noise \( \widetilde{T} \):
\begin{equation}
\label{july31-1}
P(\sigma ,\widetilde{T})=\left\{ \begin{array}{cc}
\frac{\sigma }{\sigma _{c}}\exp \left[ -\frac{1}{\widetilde{T}}\left( \frac{\sigma _{c}}{\sigma }-1\right) \right] , & 0\leq \sigma \leq \sigma _{c}\\
1, & \sigma >\sigma _{c}
\end{array}\right\} .
\end{equation}

\noindent As one can see, each fiber now has got a non-vanishing probability
\( P(\sigma ,\widetilde{T}) \) to fail under a stress \( \sigma <\sigma _{c} \)
at any non-zero noise parameter \( \widetilde{T} \). It may be noted
that {[}unlike \( T \) in (2) or (4){]} \( \widetilde{T} \) is a
dimensionless noise parameter. \( P(\sigma ,\widetilde{T}) \) increases
as \( \widetilde{T} \) increases and for \( \sigma \geq \sigma _{c} \),
\( P(\sigma ,\widetilde{T})=1 \). Unlike at \( \widetilde{T}=0 \),
the bundle therefore fails at \( \sigma <\sigma _{c} \) after a finite
time \( \tau  \). Here we assume each fiber to have a fixed threshold
\( \sigma _{c} \), while their breaking probability at any \( \sigma  \)
(\( <\sigma _{c} \)) is due to noise-activated hopping over the barrier
height (\( \sigma _{c}-\sigma  \)). This differs from the earlier
model studies \cite{Roux-00,coleman-58} where the load distribution
is noise induced.

\vskip .1 in

\noindent \textbf{(a) Failure time }

\noindent \vskip .05 in

\noindent At \( \widetilde{T}\neq 0 \) and under any stress \( \sigma  \)
(\( <\sigma _{c} \)), some fibers fail due to noise and the load
gets shared among the surviving fibers, which in turn enhances their
stress value, inducing further failure. Denoting the fraction of fibers
that remain intact at time \( t \) by \( U_{t} \), a discrete time
recursion relation (see \cite{fiber-dynamic}) can be written as \begin{equation}
\label{july25-1}
U_{t+1}=U_{t}\left[ 1-P\left( \frac{\sigma }{U_{t}},\widetilde{T}\right) \right] ,
\end{equation}

\noindent where \( \sigma /U_{t}=L/(NU_{t}) \) is the redistributed
load per fiber among the \( NU_{t} \) surviving fibers at time \( t \).
In the continuum limit, we can write the above recursion relation
in a differential form \begin{equation}
\label{july25-2}
-\frac{dU}{dt}=\frac{\sigma }{\sigma _{c}}\exp \left[ -\frac{1}{\widetilde{T}}\left( \frac{\sigma _{c}}{\sigma }U-1\right) \right] ,
\end{equation}

\noindent giving \begin{equation}
\label{july 25-3}
\tau =\int _{0}^{\tau }dt=\frac{\sigma _{c}}{\sigma }\exp \left( -\frac{1}{\widetilde{T}}\right) \int _{0}^{1}\exp \left[ \frac{1}{\widetilde{T}}\left( \frac{\sigma _{c}}{\sigma }\right) U\right] dU
\end{equation}

\noindent or \begin{equation}
\label{july 25-4}
\tau =\widetilde{T}\exp \left( -\frac{1}{\widetilde{T}}\right) \left[ \exp \left( \frac{\sigma _{c}}{\sigma \widetilde{T}}\right) -1\right] ,
\end{equation}

\noindent for \( \sigma <\sigma _{c} \). For \( \sigma \geq \sigma _{c} \),
starting from \( U_{t}=1 \) at \( t=0 \), one gets \( U_{t+1}=0 \)
from (5) and (6), giving \( \tau =0 \). 

For small \( \widetilde{T} \) and as \( \sigma \rightarrow \sigma _{c} \),
\( \tau \simeq \widetilde{T}\exp \left[ \left( \sigma _{c}/\sigma -1\right) /\widetilde{T}\right]  \).
This failure time \( \tau  \) therefore approaches infinity as \( \widetilde{T}\rightarrow 0 \).
For \( \sigma <\sigma _{c} \), one gets finite failure time \( \tau  \)
which decreases exponentially as \( \sigma  \) approaches \( \sigma _{c} \)
or as \( \widetilde{T} \) increases and \( \tau =0 \) for \( \sigma \geq \sigma _{c} \).
This last feature is absent in the earlier formulations \cite{Roux-00}.
However, all these features are very desirable and are in qualitative
agreement with the recent experimental observations \cite{expt-01}.
This is also comparable with the phenomenological results from Griffith
theory discussed in the earlier section, although the crack size effect
in the Griffith theory differs from that in the fiber bundle case.
Our numerical study confirms the above analytic results {[}obtained
using the continuum version of the recursion relation (6){]} (see
Fig. 1) well. 

\resizebox*{8cm}{6cm}{\rotatebox{-90}{\includegraphics{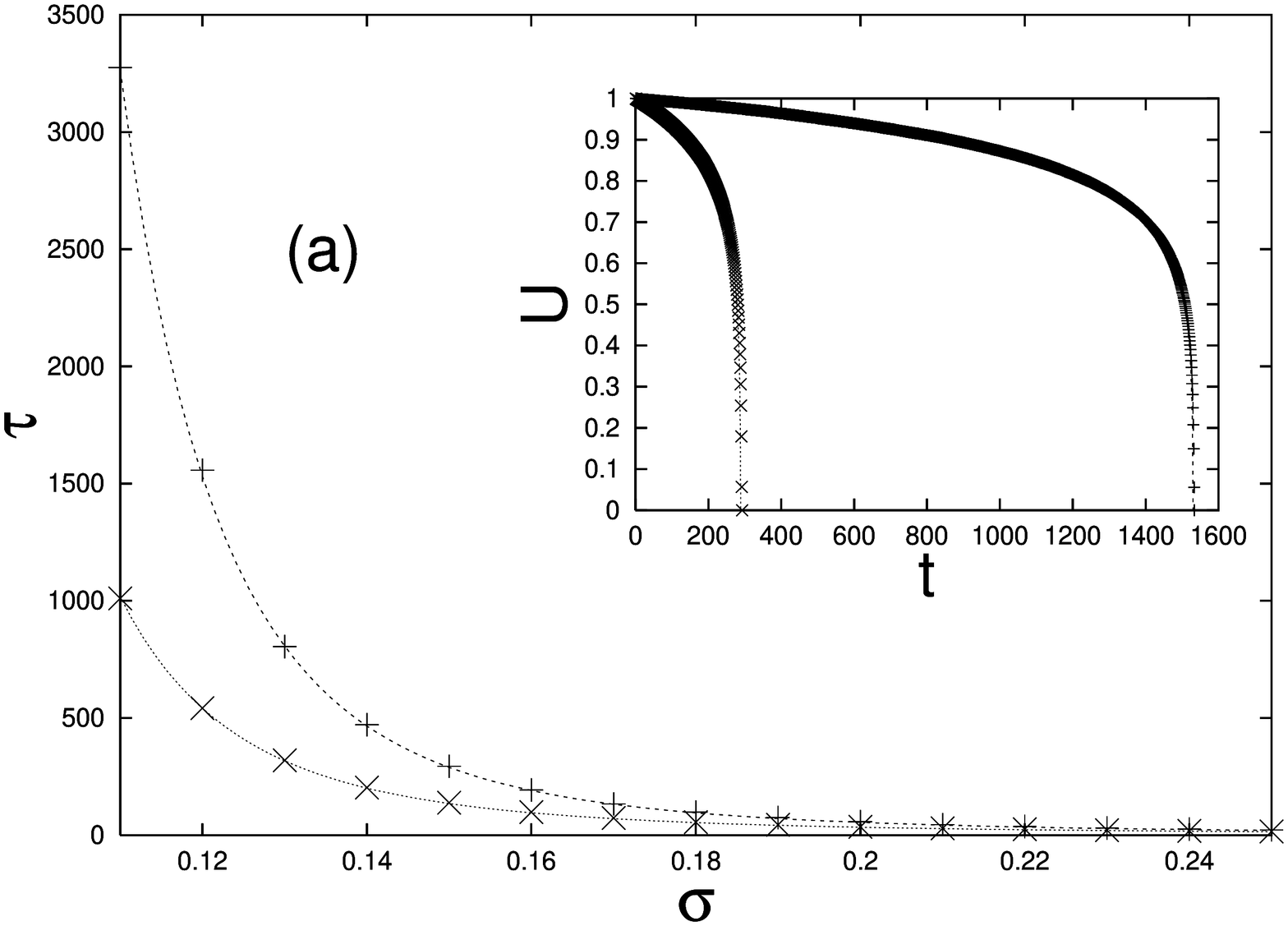}}} \resizebox*{8cm}{6cm}{\rotatebox{-90}{\includegraphics{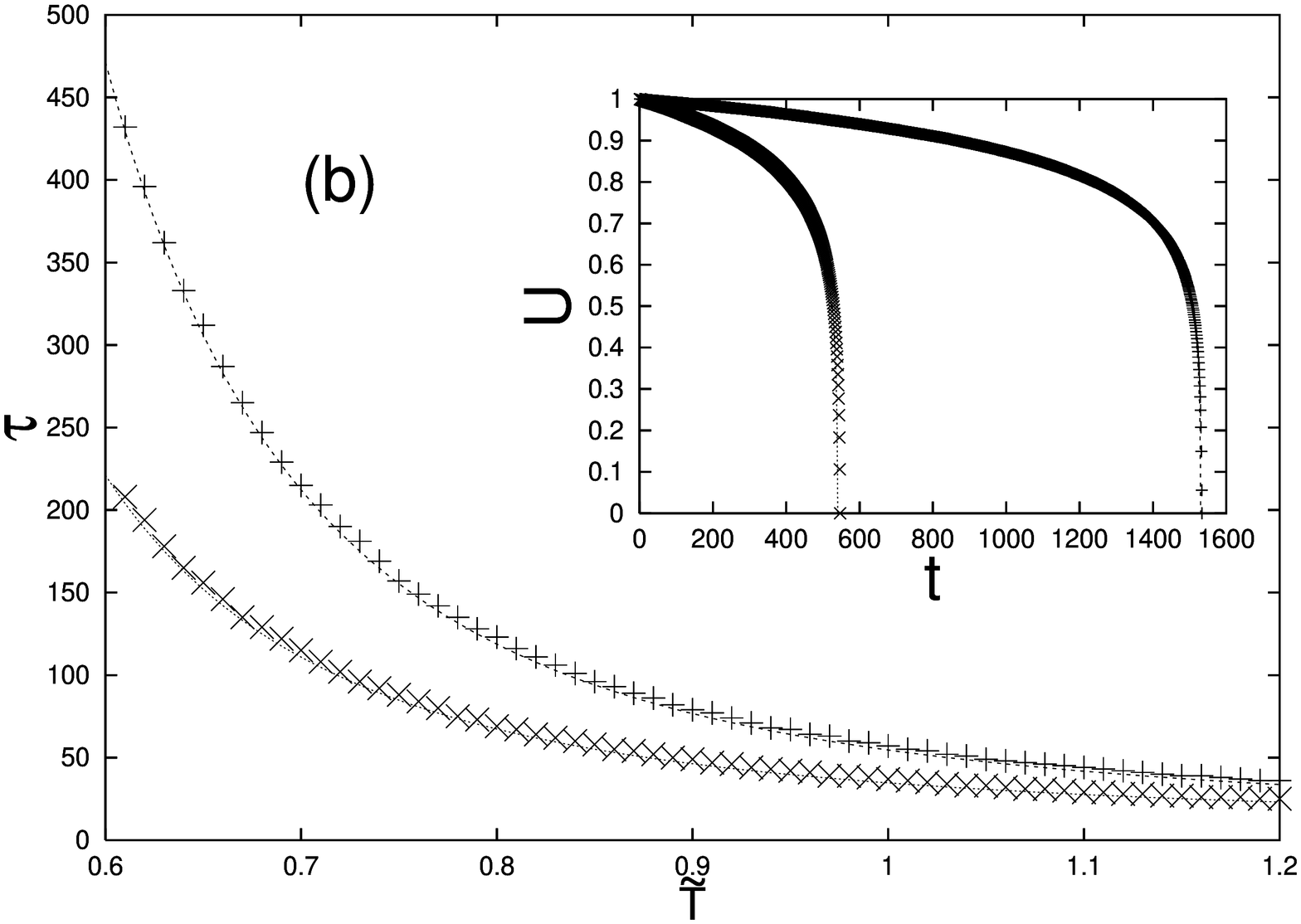}}} 

\vskip .15in

\noindent {\footnotesize Fig. 1. The simulation results showing variation
of average failure time \( \tau  \) against (a) stress \( \sigma  \)
and (b) against noise \( \widetilde{T} \), for a bundle containing
\( N=10^{5} \) fibers. The theoretical results are shown by dotted
and dashed lines {[}from eqn. (9){]}. The insets show the simulation
results for the variation of the fraction \( U \) of unbroken fibers
with time \( t \) for different \( \widetilde{T} \) values {[}1.2
(cross) and 1.0 (plus){]} in (a) and \( \sigma  \) values {[}0.15
(cross) and 0.12 (plus){]} in (b). The dotted and dashed lines represent
the theoretical results {[}eqns. (10 \& 9){]}. }{\footnotesize \par}

\vskip .2in

\noindent \textbf{(b) Avalanche size distribution }

\noindent \vskip .1in

\noindent From the recursion relations (6) or (7), one can see that
in each unit time interval a number of fibers break giving an avalanche
size for the breaking. The avalanche size therefore is given by \( dU/dt \)
and during the entire failure period \( \tau  \), different sizes
of avalanches take place. Solving for \( U(t) \) from (7) one gets
\begin{equation}
\label{july29-1}
U(t)=\frac{\sigma \widetilde{T}}{\sigma _{c}}\ln \left[ \frac{\tau -t}{\widetilde{T}\exp (-1/\widetilde{T})}+1\right] ,
\end{equation}

\noindent employing the expression (9) for \( \tau  \). One can easily
check that \( U(t)=1 \) at \( t=0 \) and \( U(t)=0 \) at \( t=\tau  \)
(see Fig. 1). Also as \( t\rightarrow t_{c}\equiv \tau  \), \( U(t) \)
decays as \( \ln (\tau -t)\sim (\tau -t)^{\beta } \) with \( \beta =0_{+} \)
from (10). Expressing \( dU/dt \) as the avalanche size \( m \),
one gets from (10) \begin{equation}
\label{july29-2}
m^{-1}\sim \frac{\tau -t}{\widetilde{T}\exp (-1/\widetilde{T})}+1\sim \tau -t,
\end{equation}

\noindent for \( \widetilde{T}\rightarrow 0 \).

Here the avalanche size \( m \) can also be interpreted as the rate
of breaking (\( dU/dt \)) and it varies with time as \( (\tau -t)^{-\gamma } \),
\( \gamma =1 \) as \( t\rightarrow t_{c}\equiv \tau  \). Since \( \tau -t \)
corresponds to the cumulative probability \( \int _{m}^{\infty }D(m)dm \)
of avalanches beyond \( t \), one gets \begin{equation}
\label{july29-3}
D(m)\sim m^{-\alpha };\alpha =2
\end{equation}

\noindent for the (differential) avalanche size distribution \( D(m) \).
Also, the exponent of power law decay in eqn. (12) is independent
of stress \( \sigma  \) and the noise level \( \widetilde{T} \)
which has been confirmed through numerical simulations (see Fig. 2). 

\vspace{0.3cm}
{\centering \resizebox*{11cm}{8cm}{\rotatebox{-90}{\includegraphics{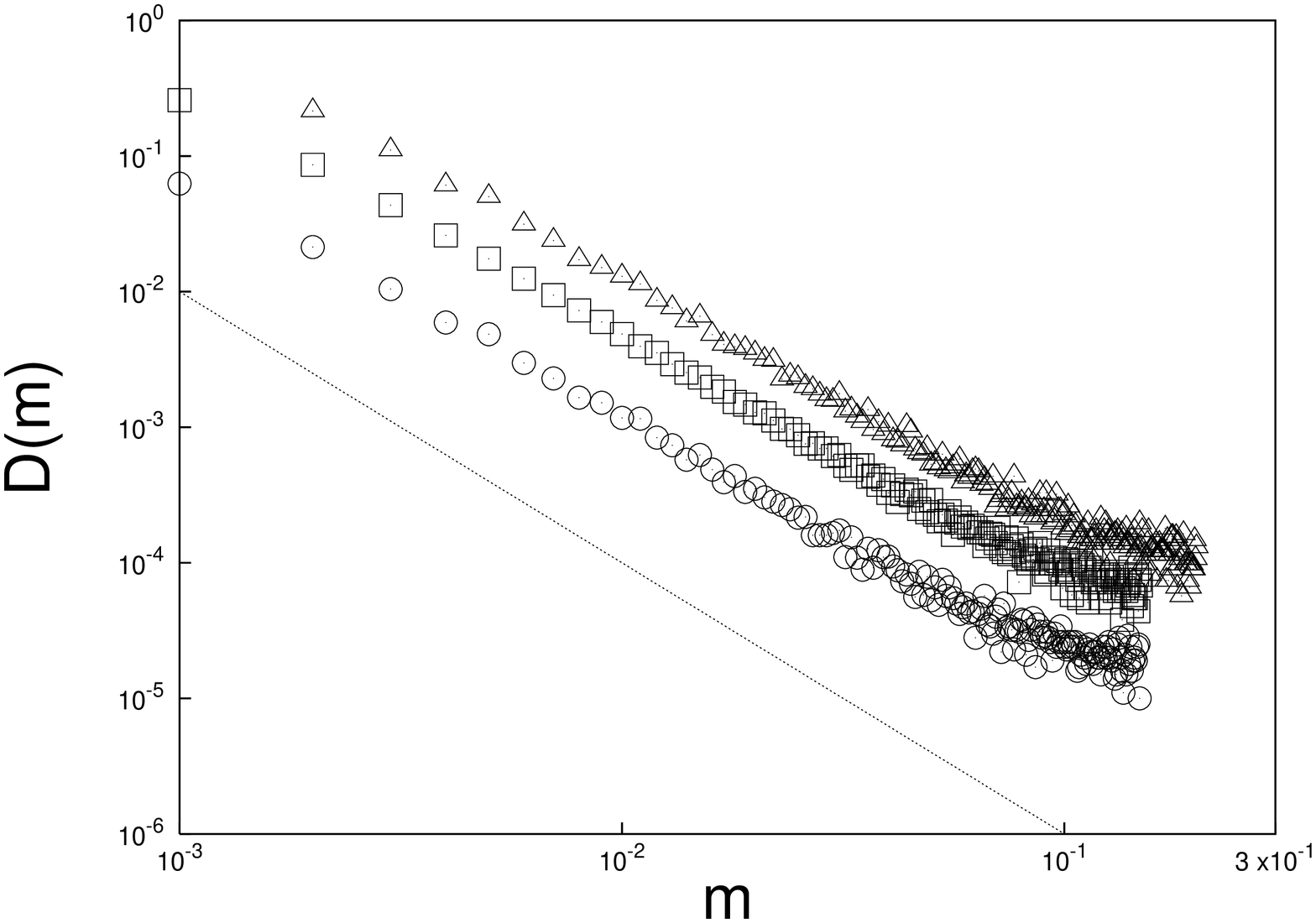}}} \par}
\vspace{0.3cm}

\vskip .1in

\noindent {\footnotesize Fig. 2. The simulation results for the distribution
\( D(m) \) of avalanches in the bundle with \( N=10^{5} \) (averaged
over \( 10^{3} \) realisations): \( \sigma =0.2, \)} \( \widetilde{T}=0.8 \)
(triangle), {\footnotesize \( \sigma =0.15, \)} \( \widetilde{T}=0.8 \)
(circle) {\footnotesize and \( \sigma =0.15, \)} \( \widetilde{T}=1.0 \)
{\footnotesize (square). The dashed line corresponds to a decay power
\( 2.0 \). }{\footnotesize \par}

\vskip .2in

\noindent It may be mentioned that such avalanches manifest in the
ultrasonic emissions during the propagation of fracture in the solid
and the ultrasonic amplitudes are also observed to have similar power
law distribution \cite{Books-1}.

\noindent \newpage

\noindent \textbf{IV. Simulation results for fatigue-failure in random
fiber bundles}

\vskip .1in

\noindent In order to investigate the fatigue behavior in random fiber
bundles we consider three different kinds of fiber strength distributions
\( \rho (\sigma _{c}) \): (A) Uniform distribution of fiber strength
where \( \rho (\sigma _{c})=1 \) for \( 0<\sigma _{c}\leq 1 \) and
\( \rho (\sigma _{c})=0 \) for \( \sigma _{c}>1 \), (B) Linearly
increasing distribution of fiber strength where \( \rho (\sigma _{c})=2\sigma _{c} \)
for \( 0<\sigma _{c}\leq 1 \) and \( \rho (\sigma _{c})=0 \) for
\( \sigma _{c}>1 \) and (C) Linearly decreasing distribution of fiber
strength where \( \rho (\sigma _{c})=2(1-\sigma _{c}) \) for \( 0<\sigma _{c}\leq 1 \)
and \( \rho (\sigma _{c})=0 \) for \( \sigma _{c}>1 \). It has been
already shown analytically \cite{fiber-dynamic}, from the dynamics
of failure in all these three kinds of fiber bundles in the absence
of any noise (vanishing \( T \) or \( \widetilde{T} \) in (5)),
the bundle's strength \( \widetilde{\sigma _{c}}=1/4 \) for model
A, \( \widetilde{\sigma _{c}}=\sqrt{4/27} \) for model B and \( \widetilde{\sigma _{c}}=4/27 \)
for model C. We now consider the effect of the noise \( \widetilde{T} \)
inducing the failure probability \( P(\sigma ,\widetilde{T})=\exp \left[ -\frac{1}{\widetilde{T}}\left( \frac{\sigma _{c}}{\sigma }-1\right) \right]  \)
for \( 0<\sigma \leq \sigma _{c} \) and \( 1 \) for \( \sigma >\sigma _{c} \),
in the (fatigue) dynamics of such bundles, where \( \sigma _{c} \)
is the strength of the individual fibers in the bundle. 

We have studied these numerically, using Monte Carlo method (for bundles
having \( N=10^{5} \) or more fibers). We have considered bundles
having the above three kinds (A, B and C) of \( \rho (\sigma _{c}) \)
one by one. The noise induced failure mentioned above is realised
only in a Monte Carlo way. Taking averages typically over \( 10^{3} \)
Monte Carlo runs the fraction of unbroken fibers \( U(t) \) at any
time \( t \) at a fixed stress level \( \sigma (<\widetilde{\sigma _{c}}) \)
is noted. At any \( \sigma  \), the average failure time \( \tau  \)
(when \( U(t)=0 \)) is extracted. The form of the distributions and
the variations of average time with noise \( \widetilde{T} \) and
stress \( \sigma  \) are shown for the three types of bundles. We
find that \( \tau  \) fits a form \begin{equation}
\label{dec4}
\tau =\widetilde{T}\exp \left( -\frac{1}{\widetilde{T}}\right) \left[ \exp \left( \frac{\widetilde{\sigma _{c}}}{\sigma \widetilde{T}}+\frac{1}{\widetilde{T}}\right) -1\right] 
\end{equation}

\noindent for all types of bundles (indicated by dotted lines in Fig.
3). We find that this phenomenological form (13) is indeed very close
to the analytic result (9) for the fixed strength fiber bundle; it
is somewhat approximate for these bundles and fits better for lower
noise (\( \widetilde{T} \)) and stress (\( \sigma  \)) levels. The
avalanche size distributions in all these three models (A, B and C)
have been studied numerically (see Fig. 4) and we find them to follow
the same power law decay (12) with \( \alpha \simeq 2.0 \). 

{\centering \resizebox*{5.2cm}{6cm}{\rotatebox{-90}{\includegraphics{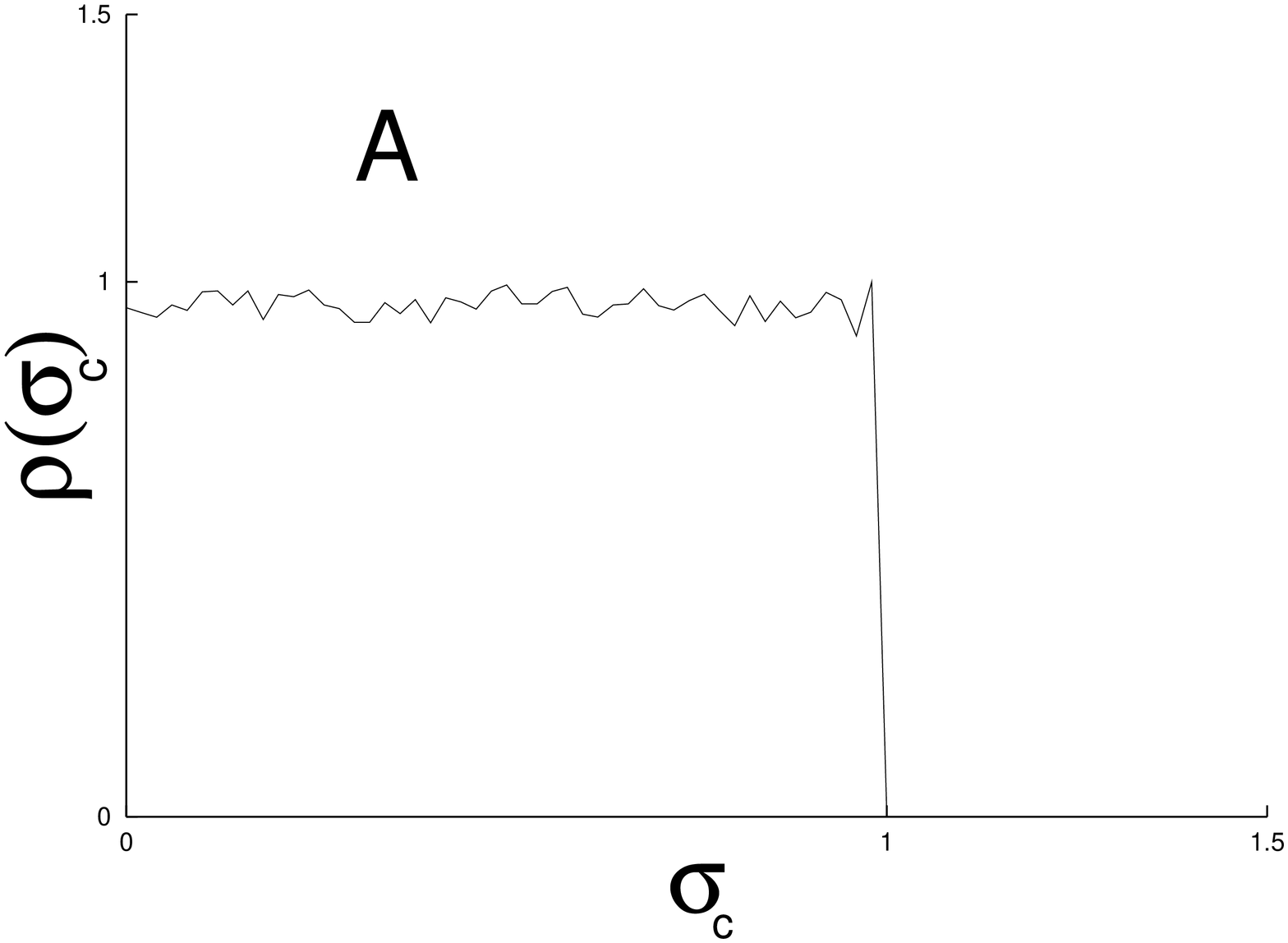}}} \resizebox*{5.2cm}{6cm}{\rotatebox{-90}{\includegraphics{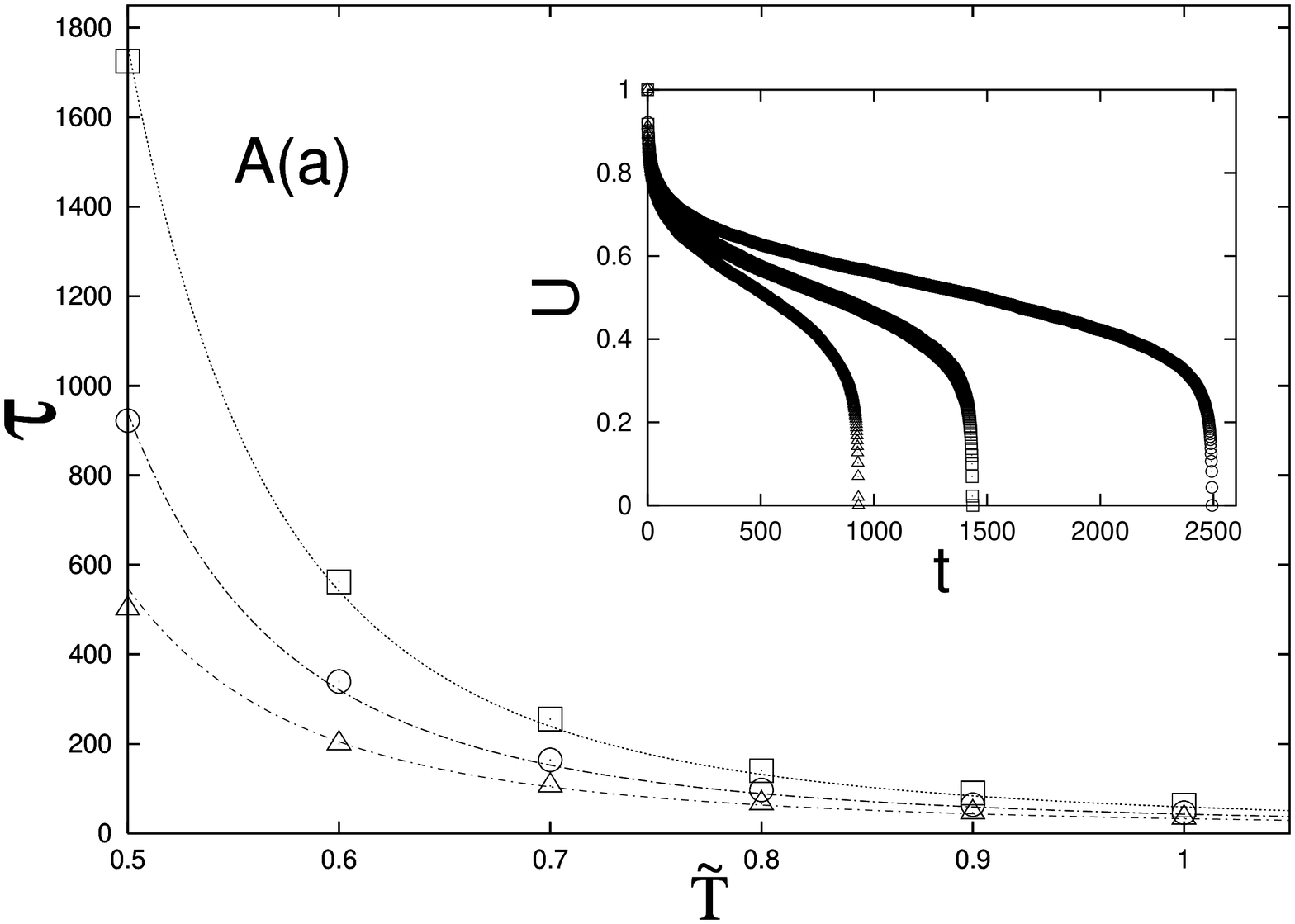}}} 
\resizebox*{5.2cm}{6cm}{\rotatebox{-90}{\includegraphics{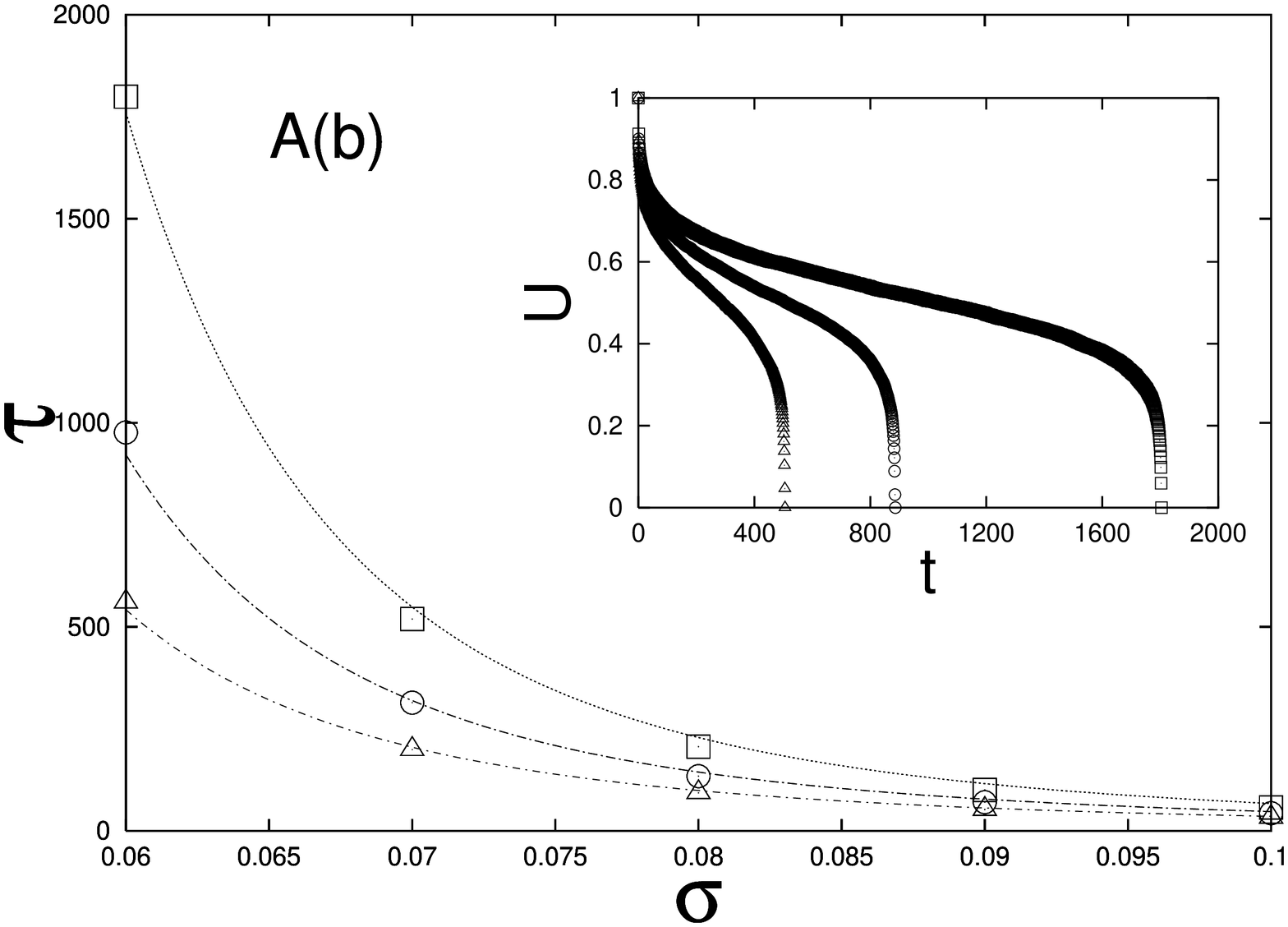}}} \par}

\resizebox*{5.2cm}{6cm}{\rotatebox{-90}{\includegraphics{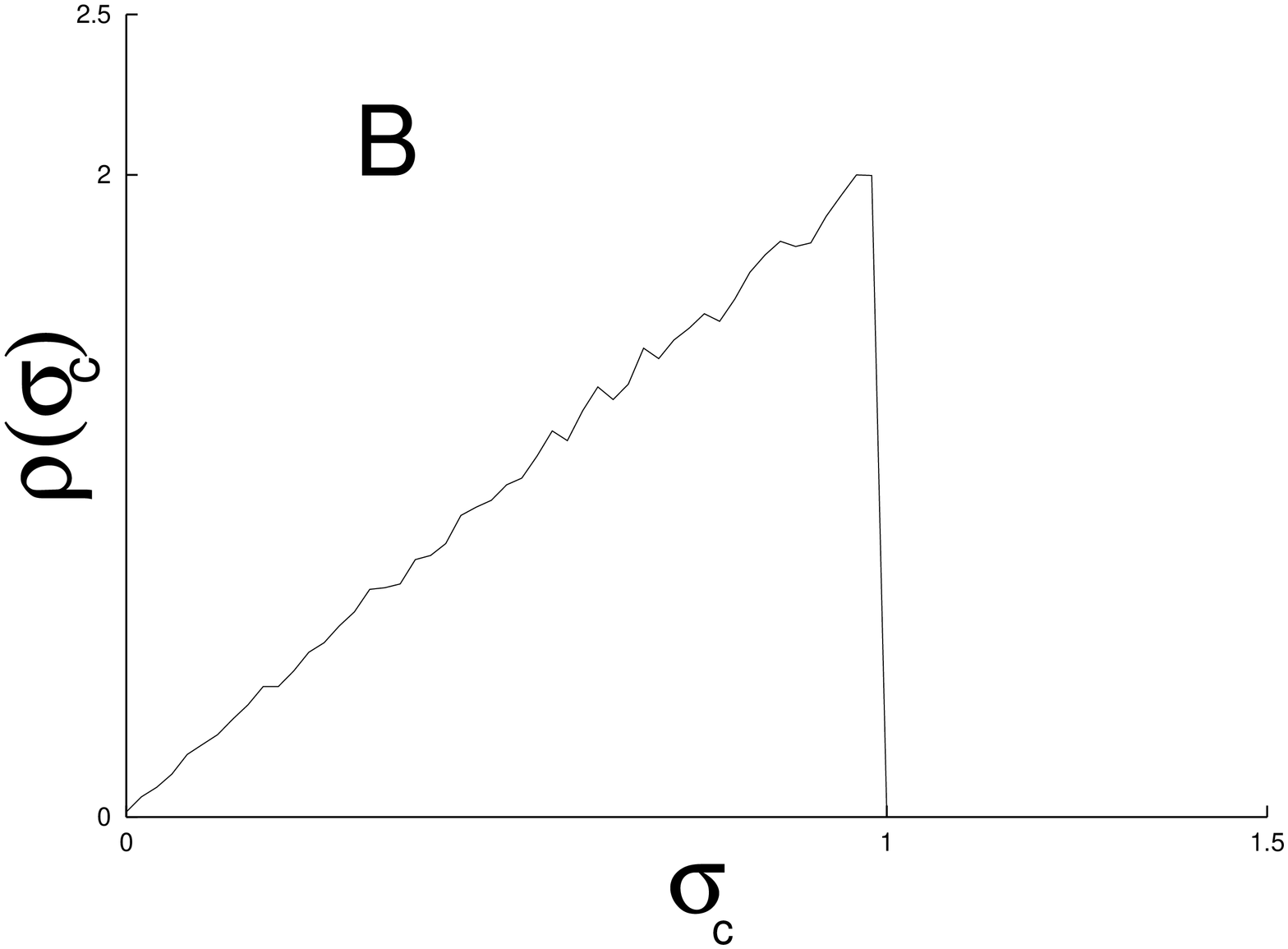}}} \resizebox*{5.2cm}{6cm}{\rotatebox{-90}{\includegraphics{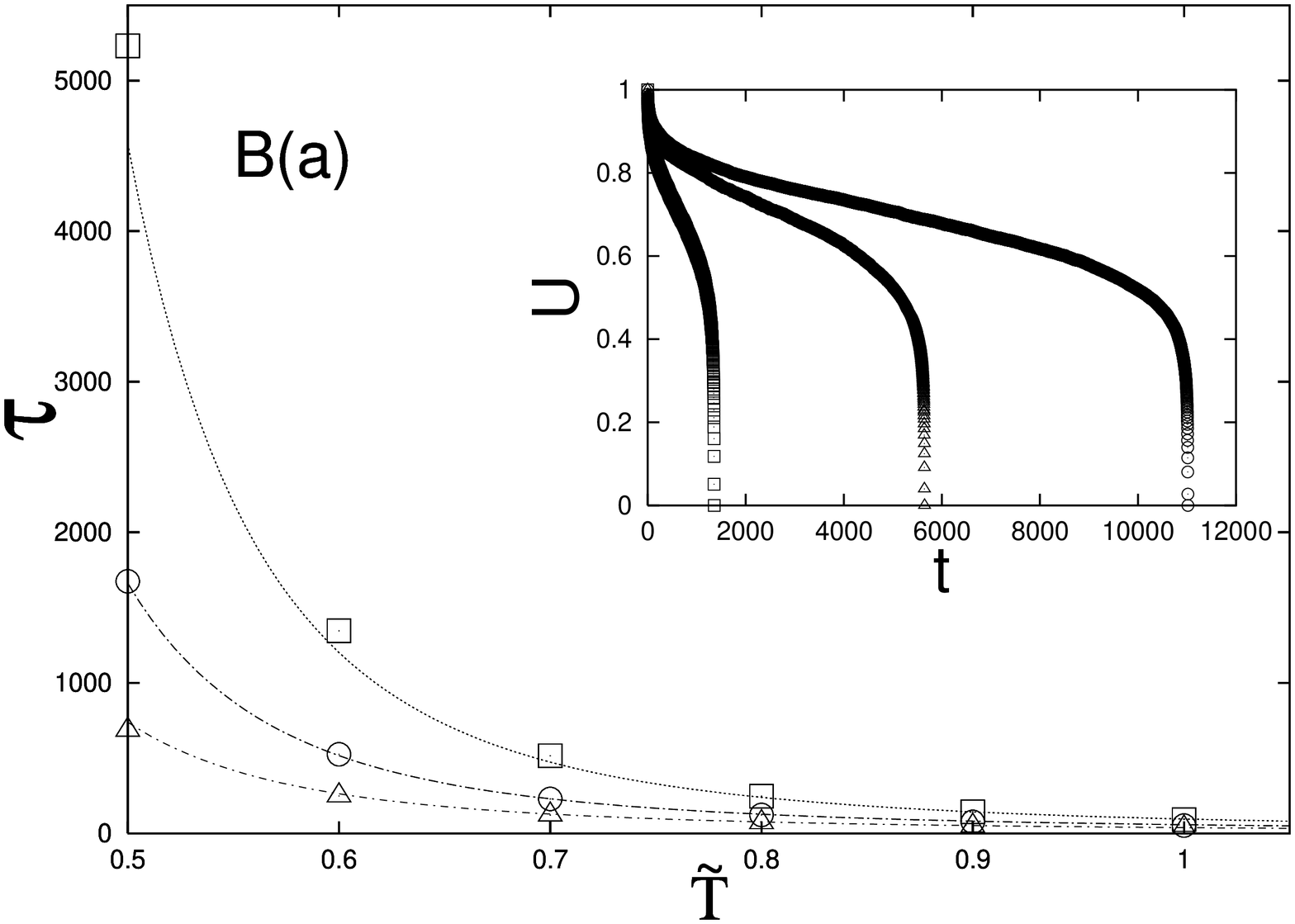}}} 
\resizebox*{5.2cm}{6cm}{\rotatebox{-90}{\includegraphics{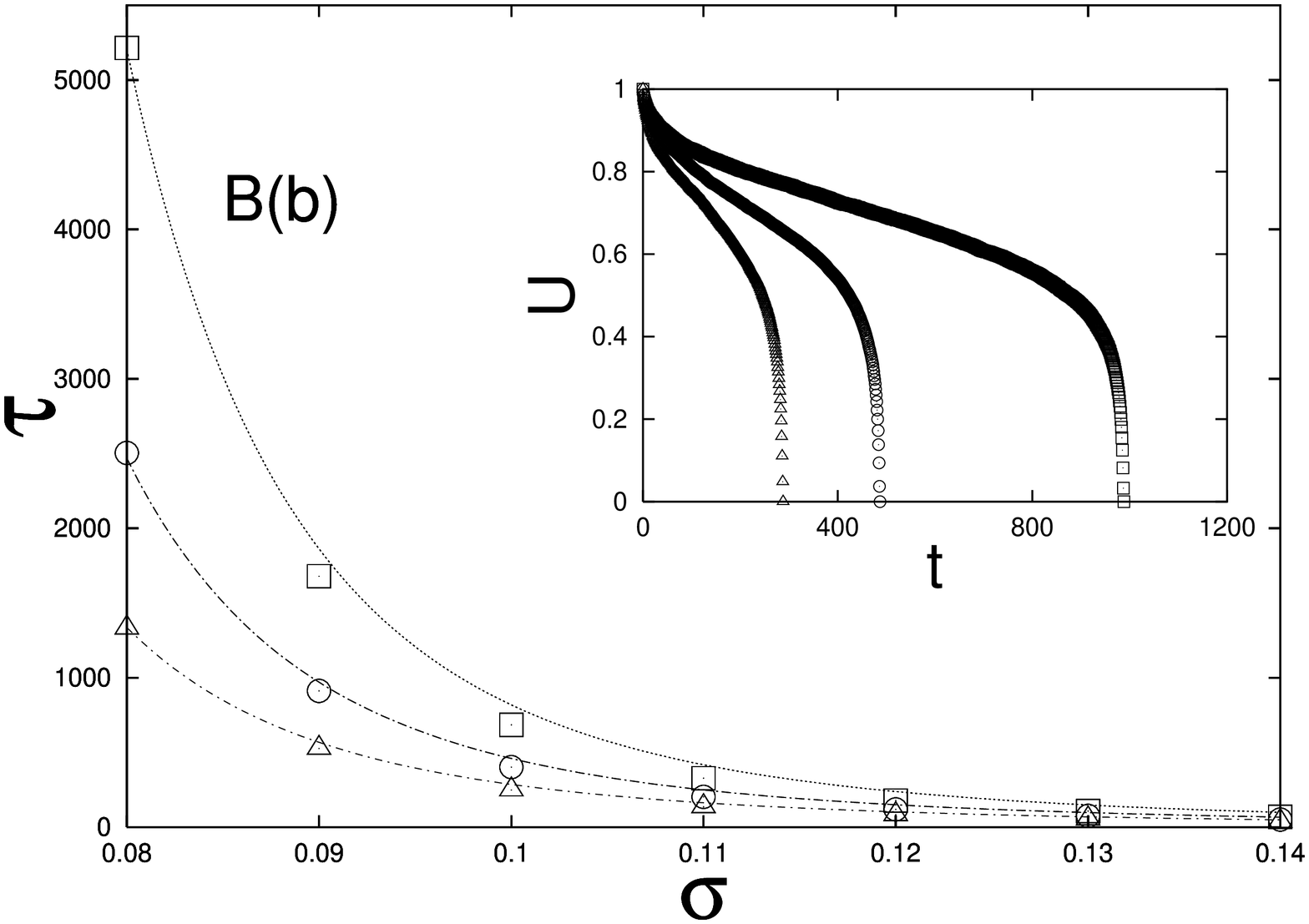}}} 

\resizebox*{5.2cm}{6cm}{\rotatebox{-90}{\includegraphics{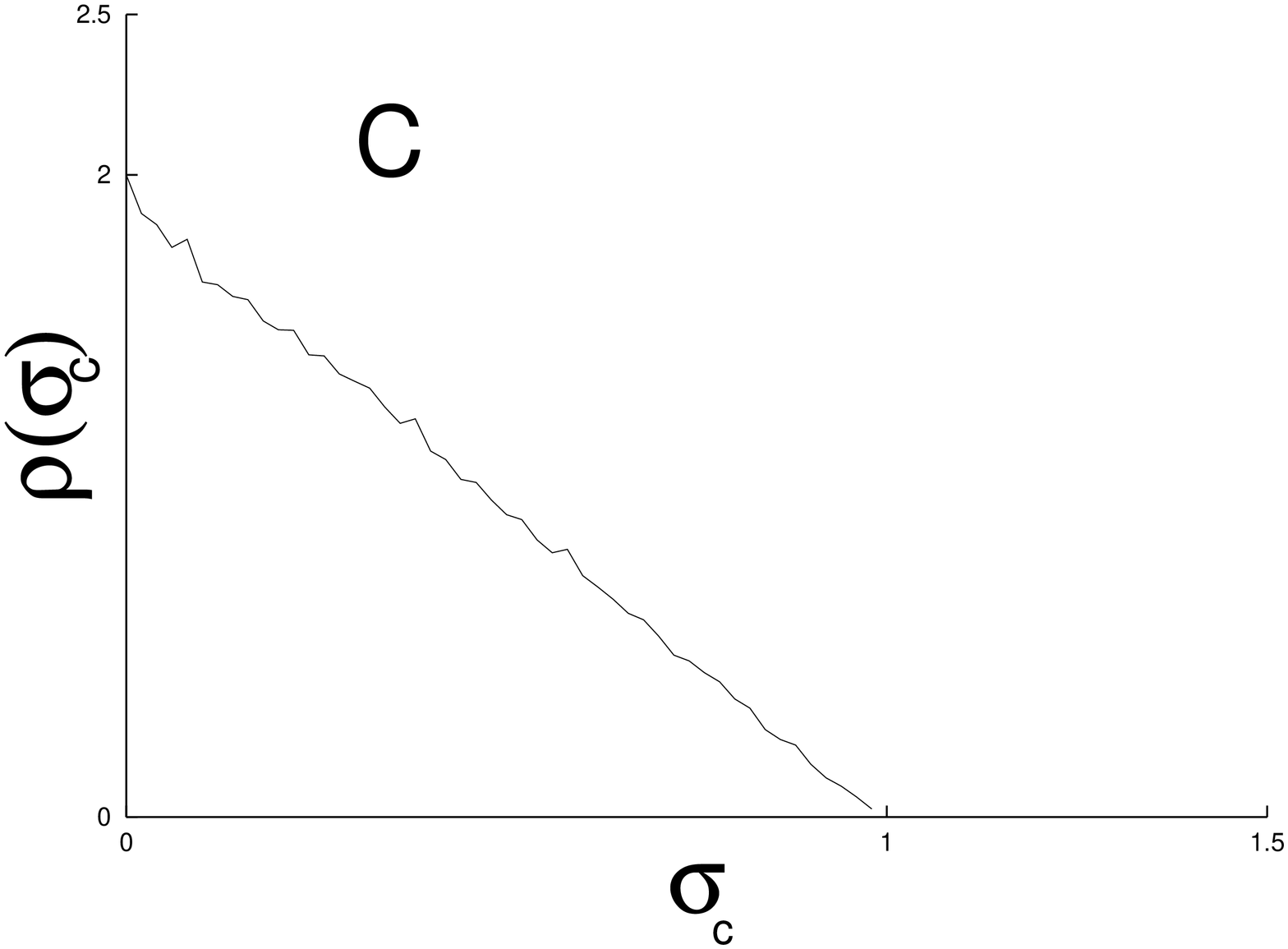}}} \resizebox*{5.2cm}{6cm}{\rotatebox{-90}{\includegraphics{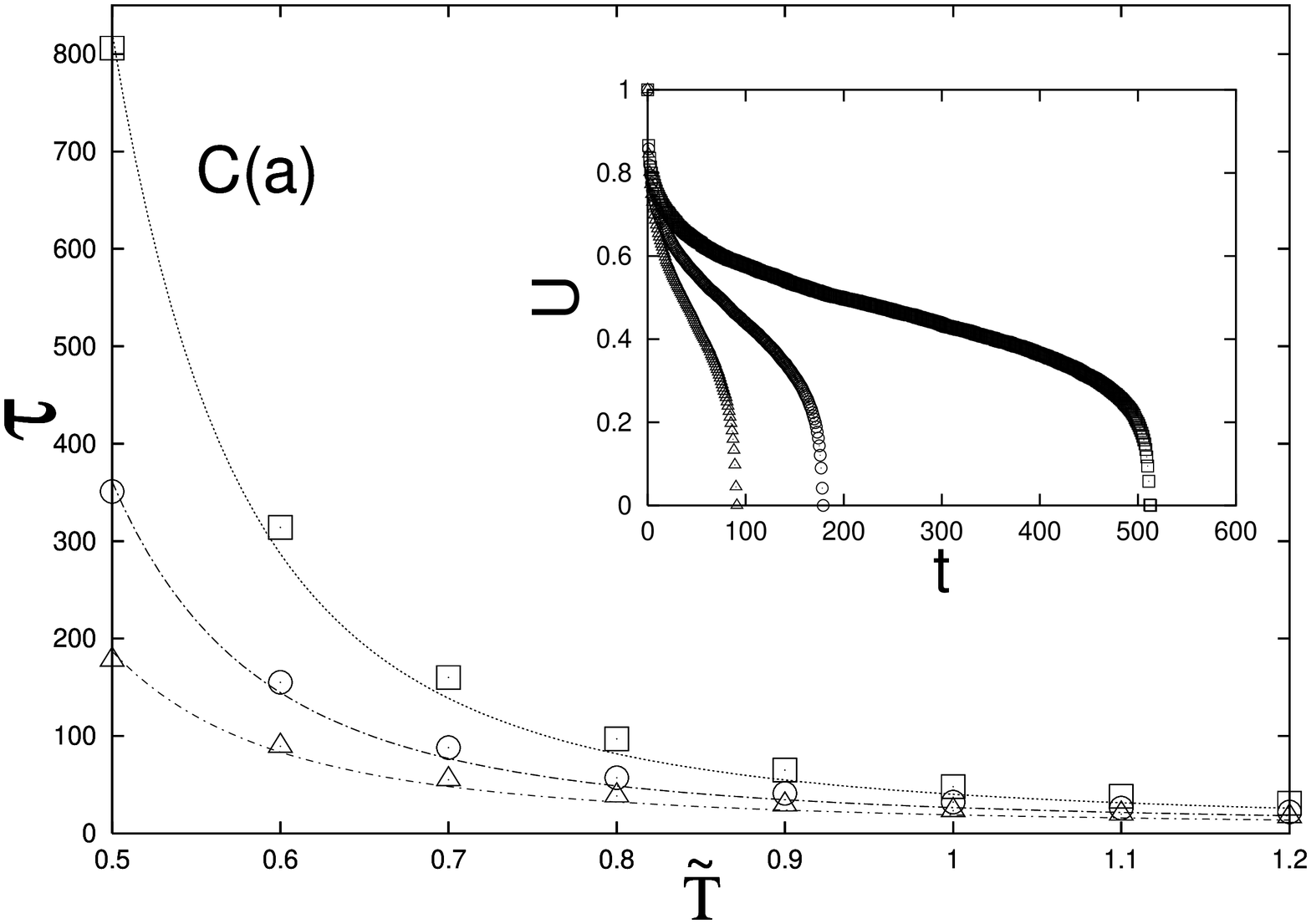}}} \resizebox*{5.2cm}{6cm}{\rotatebox{-90}{\includegraphics{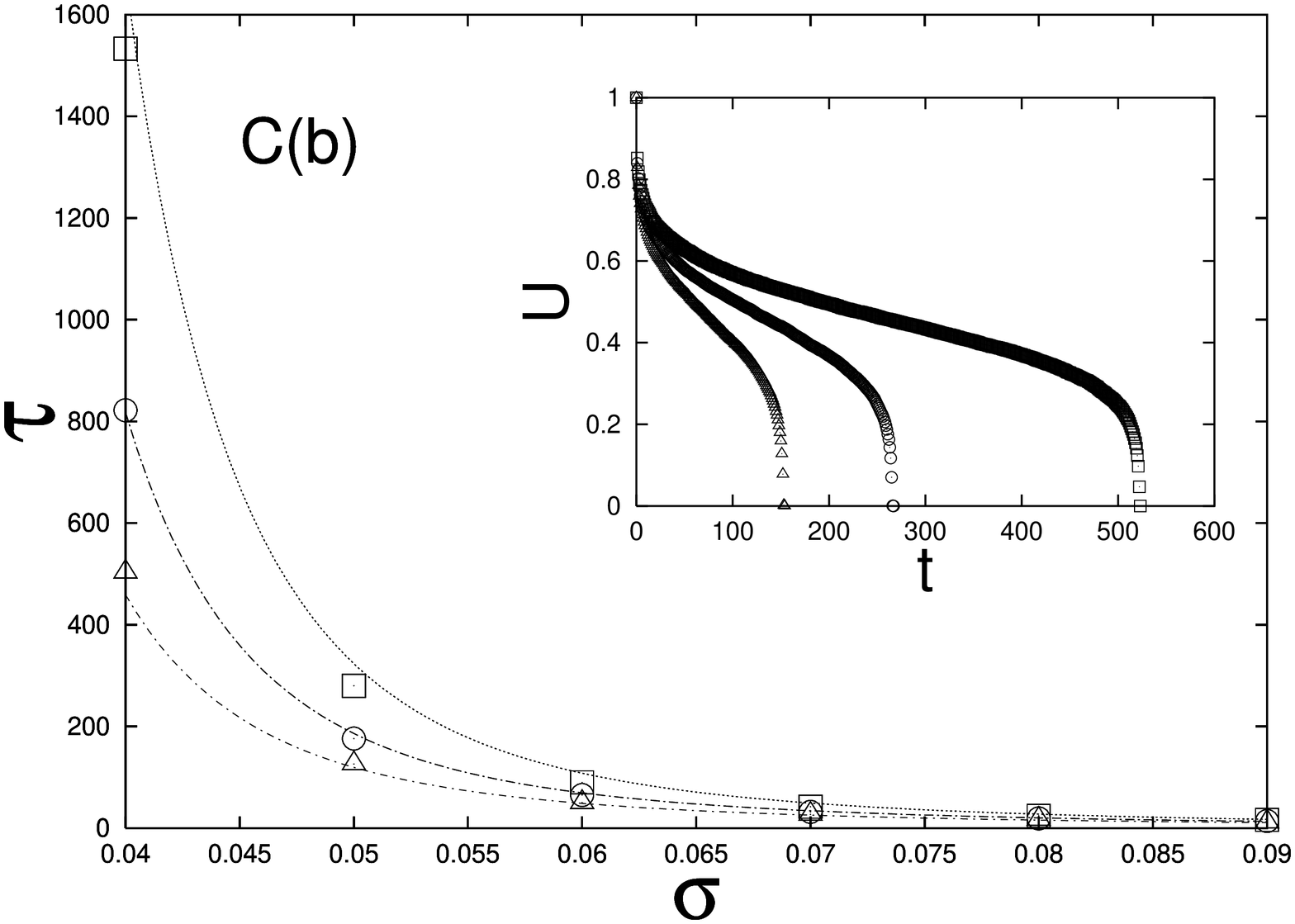}}} 

\vskip .3in

\noindent {\footnotesize Fig. 3. Typical fiber strength distributions
\( \rho (\sigma _{c}) \) considered and the simulation results for
fatigue behavior: (a) average failure time \( \tau  \) vs. noise
\( \widetilde{T} \) (for three different stress values \( \sigma  \))
and (b) \( \tau  \) vs. \( \sigma  \) (for three different noise
values \( \widetilde{T} \)) are shown for \( N=10^{5} \) fibers.
The time variation of fraction of surviving fibers are shown in the
insets for the three models: (A) with uniform \( \rho (\sigma _{c}) \),
(B) with linearly increasing \( \rho (\sigma _{c}) \) and (C) with
linearly decreasing \( \rho (\sigma _{c}) \); all having a cut off
at \( \sigma _{c}=1 \). The dotted lines in (a) and (b) corresponds
to the fit with expression (13) where \( \widetilde{\sigma _{c}}\simeq 0.245 \)
in (A) (exact value=\( 1/4 \) {[}3{]}), \( \widetilde{\sigma _{c}}\simeq 0.370 \)
in (B) (exact value=\( \sqrt{4/27} \) {[}3{]}), \( \widetilde{\sigma _{c}}\simeq 0.148 \)
in (A) (exact value=\( 4/27 \) {[}3{]}). }{\footnotesize \par}

\newpage

\vspace{0.3cm}
{\centering \resizebox*{11cm}{8cm}{\rotatebox{-90}{\includegraphics{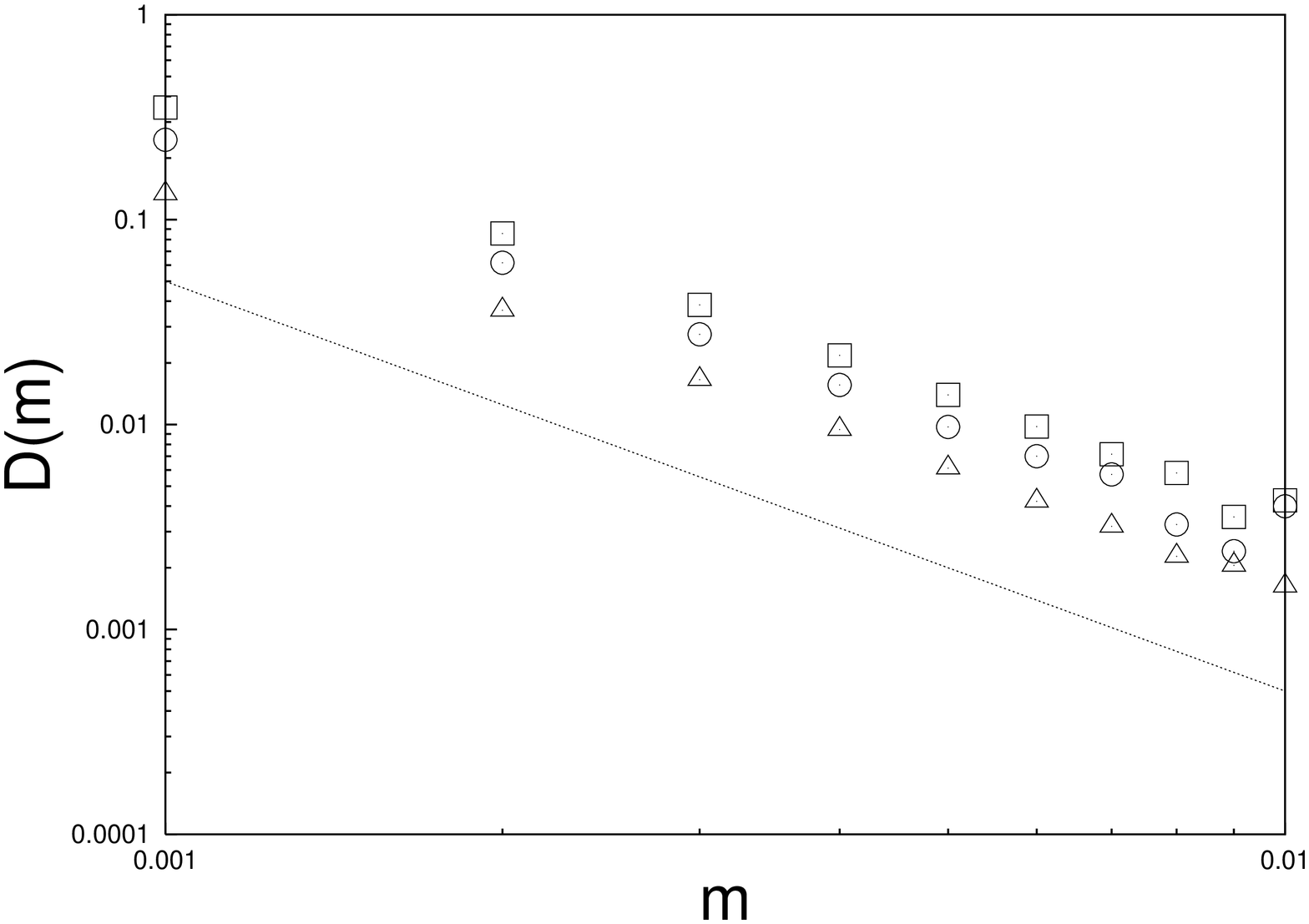}}} \par}
\vspace{0.3cm}

\noindent {\footnotesize Fig. 4. The simulation results for the distributions
\( D(m) \) of avalanches (\( m \)) in the three random fiber bundles
with \( N=10^{5} \) (averaged over \( 4\times 10^{3} \) realisations):
for model (A) with \( \sigma =0.07, \)} \( \widetilde{T}=0.5 \)
(square), {\footnotesize for model (B) with \( \sigma =0.12, \)}
\( \widetilde{T}=0.4 \) (circle) {\footnotesize and for model (C)
with \( \sigma =0.04, \)} \( \widetilde{T}=0.5 \) {\footnotesize (triangle).
The dashed line corresponds to a decay power \( 2.0 \).}{\footnotesize \par}

\vskip .2in

\noindent \textbf{V. Summary and discussions }

\noindent \vskip .1in

\noindent First, we have studied analytically the macroscopic failure
of a homogeneous fiber bundle model where each fiber has an unique
strength (\( \sigma _{c} \)). At zero noise (\( \widetilde{T}=0 \))
all the fibers of the bundle fail simultaneously for \( \sigma \geq \sigma _{c} \),
while at \( \widetilde{T}\neq 0 \) each fiber has got a non-vanishing
failure probability {[}given by eqn. (5){]} due to the thermal-like
activation. The dynamics of failure of the bundle has been solved
using the continuum version of the recursion relation (6) for global
load sharing case. The resulting expression (8) for the average failure
time (\( \tau  \)) has qualitative features similar to that (4) obtained
from the phenomenological nucleation rate theory applied for a Griffith's
crack. Both the forms have got the desirable features that \( \tau  \)
decreases exponentially as \( \sigma  \) approaches \( \sigma _{c} \)
from below and \( \tau \simeq 0 \) for \( \sigma \geq \sigma _{c} \).
As mentioned already, although the above features agree qualitatively
with the experimental observations, the precise mathematical forms
we obtained here differ from the experimentally indicated forms \cite{expt-01}.
As time \( t \) approaches \( \tau  \), the fraction of unbroken
fibers decay as \( (\tau -t)^{\beta } \), \( \beta =0_{+} \) and
its rate of breaking grows as \( (\tau -t)^{-\gamma } \), with \( \gamma =1 \).
The avalanche size distribution \( D(m) \) is also obtained analytically
for the dynamics. It is seen to have a robust power law governed decay
behavior \( D(m)\sim m^{-\alpha } \) with \( \alpha =2 \). Our numerical
results also confirm this behavior. Next, we have studied numerically
the dynamics and the average breaking time \( \tau  \) for bundles
where the breaking strengths are not fixed and are given by the three
simple distributions \( \rho (\sigma _{c}) \). We find that for all
the three cases, the average \( \tau  \) fits well a form (13), which
is very close to the analytic form for \( \tau  \) in (9) for fixed
failure threshold of the fibers. We have also investigated the avalanche
size distributions in these models and obtained the same power law
behavior, as for the fixed strength fibers. 

As mentioned already, here the noise parameter (\( \widetilde{T} \)
in (5)) can not be identified with temperature (\( T \) in (2)) which
scales with the (crack) energy. In fact, although this failure model
and its dynamics are applied here to classical breakdown phenomena
occurring in the fiber bundle model or (classical) percolating solids
\cite{Books-1}, they seem to be applicable to quantum breakdown due
to tunneling as well. Failures in quantum percolating solids beyond
their linear conducting or insulating regime, has not been studied
much (see however \cite{bkc-book}). In fact, like the fuse (or dielectric
breakdown) problems of percolating (or non-percolating) systems of
conductor-insulator networks, one can think of the field induced breakdown
of a quantum percolating system where the phase of the system is determined
through two energy scales: Fermi energy \( \epsilon _{f} \) and the
mobility edge \( \epsilon _{c} \). For \( \epsilon _{f}> \) \( \epsilon _{c} \)
the system is in conducting phase and it goes to insulating phase
for \( \epsilon _{f}< \) \( \epsilon _{c} \). This metal-insulator
transition at \( \epsilon _{f}= \) \( \epsilon _{c} \) (in higher
than two dimensional systems) and the scaling property of conductivity
for \( \epsilon _{f}> \) \( \epsilon _{c} \) have been studied extensively
\cite{quantum casees,Ramakrishnan-85}. For the insulating phase (\( \epsilon _{f}< \)
\( \epsilon _{c} \)), one can have electric field induced (Zener
type) breakdown (similar to dielectric breakdown of non-percolating
classical networks). This Zener breakdown of Anderson insulators or
the quantum tunneling induced breakdown of impure (localised) insulators
have not been studied much (see however \cite{bkc-book,Zhao-98}).
Unlike a (classical) fiber bundle model considered here, where all
the fibers are in parallel, one can consider a dielectric composed
of several elements in series having non-zero failure probability
for each element due to quantum tunneling (like the noise-induced
activation considered here). Any microscopic failure of such an element
would result in increased field on the surviving elements and this
in turn would enhance their failure probability. A similar dielectric
failure time (\( \tau  \)) in such quantum or Anderson insulators
is thus expected under electric field. Here \( \sigma  \) and \( \sigma _{c} \)
would be replaced by \( \epsilon _{f} \) and \( \epsilon _{c} \)
respectively and \( \widetilde{T} \) would correspond to the inverse
tunneling length determined by the electric field (with the Planck's
constant as the proportionality factor, incorporating the intrinsic
noise) \cite{bkc-book}. 

Our study here for fatigue breakdown in the model fiber bundles shows
that the average failure time for the bundle at a stress value \( \sigma  \)
less than the bundle strength \( \sigma _{c} \) (for homogeneous
fiber bundle) or \( \widetilde{\sigma _{c}} \) (for random fiber
bundles), above which the bundle fails immediately, decreases exponentially
as \( \sigma  \) approaches \( \sigma _{c} \) or \( \widetilde{\sigma _{c}} \)
from below. This has already been observed in several experiments
qualitatively. We have demonstrated this fatigue behavior here both
analytically and numerically for a homogeneous fiber bundle (section
III) and also numerically for random fiber bundles with nontrivial
strength distributions (section IV). We also believe that these observations
will be useful in quantum breakdown phenomena.

\newpage

\begin{thebibliography}{10}
\bibitem[1]{Books-1}B. K. Chakrabarti and L. G. Benguigui, \textit{Statistical Physics
of Fracture and Breakdown in Disorder Systems}, Oxford Univ. Press,
Oxford (1997); see also H. J. Herrmann and S. Roux (Eds), \emph{Statistical
Models of Disordered Media}, North Holland, Amsterdam (1990); M. Sahimi,
Phys. Rep\textbf{. 306}, 213 (1980).
\bibitem[2]{fiber-static}H. E. Daniels, Proc. R. Soc. London A \textbf{183}, 405 (1945); S.
L. Phoenix, Adv. Appl. Prob. \textbf{11}, 153 (1979); P. C. Hemmer
and A. Hansen, J. Appl. Mech. \textbf{59}, 909 (1992); R. da Silveira,
Am. J. Phys. \textbf{67}, 1177 (1999); R. C. Hidalgo, F. Kun and H.
J. Herrmann, Phys. Rev. E \textbf{64}, 066122 (2001). 
\bibitem[3]{fiber-dynamic}S. Pradhan and B. K. Chakrabarti, Phys. Rev. E \textbf{65}, 016113
(2002); S. Pradhan, P. Bhattacharyya and B. K. Chakrabarti, Phys.
Rev. E \textbf{66}, 016116 (2002); P. Bhattacharyya, S. Pradhan and
B. K. Chakrabarti (2002) to appear in Phys. Rev. E (in press); arXiv:cond-mat/0207393.
\bibitem[4]{Lawn-93}See e. g., B. R. Lawn, \emph{Fracture of Brittle Solids}, Cambridge
University Press, Cambridge (1993).
\bibitem[5]{coleman-58}B. D. Coleman, J. Appl. Phys. \textbf{29}, 968 (1958).
\bibitem[6]{Feng-91}L. Golubovic and S. Feng, Phys. Rev. A \textbf{43}, 5223 (1991). 
\bibitem[7]{expt-01}R. Banerjee and B. K. Chakrabarti, Bull. Mater. Sci. \textbf{24},
161 (2001); A. Guarino, S. Ciliberto, A. Garcimartin, M. Zei and R.
Scorretti, cond-mat/0201257 (2002).
\bibitem[8]{bkc-book}B. K. Chakrabarti in \emph{Nonlinearity and Breakdown in Soft Condensed
Matte}r, (Eds) K. K. Bardhan, B. K. Chakrabarti and A. Hansen, Springer-Verlag,
Heidelberg (1994), pp 171-185.
\bibitem[9]{Roux-00}S. Roux, Phys. Rev. E \textbf{62}, 6164 (2000); R. Scorretti, S. Ciliberto
and A. Guarino, Europhys. Lett. \textbf{55}, 626 (2001). 
\bibitem[10]{quantum casees}T. Odagaki and K. C. Chang, Phys. Rev. B \textbf{30}, 1612 (1984);
I. Chang, Z. Lev, A. B. Harris, J. Adler and A. Aharony, Phys. Rev.
Lett. \textbf{74}, 2094 (1995); I. Kh. Zharekeshev and B. Kremer,
Phys. Rev. Lett. \textbf{79}, 717 (1997); A. Kaneko and T. Ohtsuki,
J. Phys. Soc. Jap. \textbf{68}, 1488 (1999).
\bibitem[11]{Ramakrishnan-85}P. Lee and T. V. Ramakrishnan, Rev. Mod. Phys. \textbf{57}, 287 (1985).
\bibitem[12]{Zhao-98}J. Rotvig, A. P. Jauho and H. Smith, Phys. Rev. B \textbf{54}, 17691
(1996); X. -G. Zhao, W. -X. Yan and D. W. Hone, Phys. Rev. B \textbf{57},
9849 (1998).
\end{thebibliography}
\end{document}